\documentclass[a4paper,11pt,twoside,fleqn]{article}
\usepackage{amsfonts,amstext,latexsym,times,xspace}
\usepackage{epsfig}

\newcommand{\lb}[1]{\label{#1}}
\newcommand{\Eq}[1]{(\ref{#1})}
\renewcommand{\[}{\begin{eqnarray}}
\renewcommand{\]}{\end{eqnarray}}
\newcommand{\nn}{\nonumber}
\newcommand{\non}{\nonumber \\ }
\renewcommand{\=}{\equiv}
\def\ba{\begin{array}}
\def\ea{\end{array}}
\makeatletter
\newif\if@fewtab\@fewtabtrue
\makeatother

\def\moth{\mathsurround=0pt}
\newdimen\zo \zo=0pt

\def\tick{\leaders\hrule height 0.5ex depth 0pt \hskip 0.5pt}
\def\upboxfill{$\moth \setbox\zo\hbox{\tick}%
  \hskip 2pt\hbox to 0pt{$\tick$\hss}\hrulefill \hbox to 2pt{$\tick$\hss}$}
\def\underbox#1{\offinterlineskip{\mathord{\mathop{\vtop{\moth\ialign{##\crcr
      $\hfil\displaystyle{#1}\hfil$\crcr\noalign{}
      {\upboxfill}\crcr\noalign{}}}}\limits}}}
\def\dtick{\leaders\hrule height .34pt depth 0.5ex \hskip 0.5pt}
\def\downboxfill{$\moth \setbox\zo\hbox{\dtick}%
  \hskip 2pt\hbox to 0pt{$\dtick$\hss}\hrulefill%
  \hbox to 2pt{$\dtick$\hss}$}
\def\overbox#1{\mathop{\vbox{\moth\ialign{##\crcr\noalign{}
\downboxfill\crcr\noalign{\vskip 1pt\nointerlineskip}
      $\hfil\displaystyle{#1}\hfil$\crcr}}}\limits}

\newcommand{\undersym}[1]{\underbox{{}#1}}
\newcommand{\oversym}[1]{\!\overbox{{}#1}}

\newcommand{\cD}{\ensuremath{\mathcal{D}}\xspace}
\newcommand{\Gtwo}[1][{}]{\ensuremath{G_{2 #1}}\xspace}
\newcommand{\4}[1][{}]{\ensuremath{F_{4 #1}}\xspace}
\newcommand{\6}[1][{}]{\ensuremath{E_{6 #1}}\xspace}
\newcommand{\7}[1][{}]{\ensuremath{E_{7 #1}}\xspace}
\newcommand{\8}[1][{}]{\ensuremath{E_{8 #1}}\xspace}
\newcommand{\E}{E_{8(8)}}
\newcommand{\EE}{E_{7(7)}}
\newcommand{\EEE}{E_{6(6)}}

\newcommand{\USp}[1][8]{\ensuremath{\mbox{USp$(#1)$}}\xspace}
\newcommand{\SU}[1][8]{\ensuremath{\mbox{SU$(#1)$}}\xspace}
\newcommand{\SO}[1][8]{\ensuremath{\mbox{SO$(#1)$}}\xspace}

\newcommand{\SLR}[1][8]{\ensuremath{\mbox{SL$(#1,\mathbb{R})$}}\xspace}

\newcommand{\fg}{\ensuremath{\mathfrak{g}}\xspace}

\newcommand{\gO}[1][{}]{\Omega^{#1}\xspace}
\newcommand{\gOd}[1][{}]{\Omega_{#1}\xspace}

\newcommand{\gd}[1][{}]{\delta_{#1}{}}

\newcommand{\eps}[1][{}]{\epsilon^{#1}\xspace}
\newcommand{\epsd}[1][{}]{\epsilon_{#1}\xspace}

\newcommand{\Zt}{\tilde{Z}{}}
\newcommand{\Zbt}{\bar{Z}{}}

\newcommand{\Et}{\tilde{E}}
\newcommand{\Ht}{\tilde{H}}
\newcommand{\Ft}{\tilde{F}}
\newcommand{\Gt}{\tilde{G}}
\newcommand{\Xt}{\tilde{X}}
\newcommand{\Yt}{\tilde{Y}}
\newcommand{\Zti}{\tilde{Z}}

\newcommand{\ft}[2]{{\textstyle {\frac{#1}{#2}} }}
\newcommand{\2}{{\textstyle {\frac{1}{2}} }}
\newcommand{\st}{^\ast}

\newcommand{\rep}[1]{\ensuremath{\mbox{\mathversion{bold}$\mathbf{#1}$%
                     \mathversion{normal}}}}

\renewcommand{\i}{\mathrm{i}}

\newcommand{\cX}{\mathcal{X}}
\newcommand{\cY}{\mathcal{Y}}

\newcommand{\cN}{\mathcal{N}}
\newcommand{\cI}{\mathcal{I}}

\newcommand{\Com}[2]{[#1\, ,\,#2]}
\newcommand{\Sympl}[2]{\left\langle #1,#2\right\rangle}
\newcommand{\FTP}[4][{}]{\left( #2,#3,#4 \right)^{#1}}
\newcommand{\JTP}[4][{}]{\{ #2\,#3\,#4 \}^{#1}}

\newcommand{\Cn}{\ensuremath{\mathbb{C}}\xspace}
\newcommand{\Rn}{\ensuremath{\mathbb{R}}\xspace}

\newcommand{\Zn}{\ensuremath{\mathbb{Z}}\xspace}

\newcommand{\Hn}{\ensuremath{\mathbb{H}}\xspace}
\newcommand{\On}{\ensuremath{\mathbb{O}}\xspace}

\begin{document}
\thispagestyle{empty}

\begin{flushright}
AEI-2000-043\\
hep-th/0008063\\
CERN-TH/2000-230
\end{flushright}
\renewcommand{\thefootnote}{\fnsymbol{footnote}}
\setcounter{footnote}{0} 

\begin{center}
\mathversion{bold}
{\bf\Large Conformal and Quasiconformal} \medskip

{\bf\Large Realizations of Exceptional Lie Groups\footnote{This work was 
supported in part by the NATO collaborative research grant CRG. 960188.}
}
\bigskip\bigskip
\mathversion{normal}

\setcounter{footnote}{2} 
{\bf\large M.~G\"unaydin~\footnote{Work supported in
    part by the National Science Foundation under grant number PHY-9802510.
    Permanent address: Physics Department, Penn State University, University
    Park, PA 16802, USA.}\medskip\\}
{\large CERN, Theory Division\\
  1211 Geneva 23, Switzerland} \smallskip\\ {\small E-mail:
  murat@phys.psu.edu\medskip} \bigskip

{\bf\large K.~Koepsell, H.~Nicolai\medskip\\ }
{\large Max-Planck-Institut f{\"u}r Gravitationsphysik,\\
Albert-Einstein-Institut,\\
M\"uhlenberg 1, D-14476 Potsdam, Germany}
\smallskip\\ {\small E-mail:
koepsell@aei.mpg.de, nicolai@aei.mpg.de} 

\end{center}
\bigskip
\medskip

\begin{abstract}
We present a nonlinear realization of $E_{8(8)}$ on a space of 
57 dimensions, which is quasiconformal in the sense that it
leaves invariant a suitably defined ``light cone'' in $\Rn^{57}$. 
This realization, which is related to the Freudenthal triple 
system associated with the unique exceptional Jordan algebra over 
the split octonions, contains previous conformal realizations 
of the lower rank exceptional Lie groups on generalized space 
times, and in particular a conformal realization of $\EE$ on
$\Rn^{27}$ which we exhibit explicitly. Possible applications
of our results to supergravity and \hbox{M-Theory} are briefly mentioned.
\end{abstract}

\vfill

\leftline{{\sc August 2000}}
\renewcommand{\thefootnote}{\arabic{footnote}}
\setcounter{footnote}{0}
\newpage

\section{Introduction}

It is an old idea to define generalized space-times by association with Jordan
algebras $J$, in such a way that the space-time is coordinatized by the
elements of $J$, and that its rotation, Lorentz, and conformal group can be
identified with the automorphism, reduced structure, and the linear fractional
group of $J$, respectively~\cite{Guna75,Guna80,Guna89}. The aesthetic 
appeal of this idea rests to a large extent on the fact that key ingredients 
for formulating relativistic quantum field theories over four dimensional
Minkowski space extend naturally to these generalized space times; in 
particular, the well-known connection between the positive energy
unitary representations of the four dimensional conformal group $SU(2,2)$
and the covariant fields transforming in finite dimensional representations 
of the Lorentz group $SL(2,\Cn)$~\cite{Mack69,Mack77} extends to all
generalized space-times defined by Jordan  algebras~\cite{Guna99}. The 
appearance of exceptional Lie groups and algebras in extended supergravities 
and their relevance to understanding the non-perturbative regime of string 
theory have provided new impetus; indeed, possible applications to string 
and \hbox{M-Theory} constitute the main motivation for the present 
investigation.

In this paper, we will present a novel construction involving the
maximally extended Lie group $E_{8(8)}$. This construction of
$E_{8(8)}$ together with the corresponding construction of
$E_{8(-24)}$  contain all previous examples of generalized
space-times based on exceptional Lie groups, and at the same time
goes beyond the framework of Jordan algebras. More precisely, we
show that there exists a quasiconformal nonlinear realization of
$E_{8(8)}$ on a space of 57 dimensions\footnote{A nonlinear
realization will be referred to as ``quasiconformal'' if it is
based on a five graded decomposition of the underlying Lie algebra
(as for $\E$); it will be called ``conformal'' if it is based on a
three graded decomposition (as e.g. for \7[(7)]).}. This space may
be viewed as the quotient of $E_{8(8)}$ by its maximal parabolic
subgroup~\cite{Jose74,Jose76}; there is no Jordan algebra directly associated
with it, but it can be related to a certain Freudenthal triple system which
itself is associated with the ``split'' exceptional Jordan algebra
$J_3^{\On_S}$ where ${\On_S}$ denote the split real form of the octonions
$\On$. It furthermore admits an $E_{7(7)}$ invariant norm form $\cN_4$, which
gets multiplied by a (coordinate dependent) factor under the nonlinearly
realized ``special conformal'' transformations.  Therefore the light cone,
defined by the condition $\cN_4 = 0$, is actually invariant under the full
$E_{8(8)}$, which thus plays the role of a generalized conformal group. By
truncation we obtain quasiconformal realizations of other exceptional Lie
groups. Furthermore, we recover previous conformal realizations of the lower
rank exceptional groups (some of which correspond to Jordan algebras).  In
particular, we give a completely explicit conformal M\"obius-like nonlinear
realization of $E_{7(7)}$ on the 27-dimensional space associated with the
exceptional Jordan algebra $J_3^{\On_S}$, with linearly realized subgroups
$F_{4(4)}$ (the ``rotation group'') and $E_{6(6)}$ (the ``Lorentz group'').
Although in part this result is implicitly contained in the existing
literature on Jordan algebras, the relevant transformations have not been
exhibited explicitly so far, and are here presented in the basis that is also
used in maximal supergravity theories.

The basic concepts are best illustrated in terms of a simple and familiar 
example, namely the conformal group in four dimensions~\cite{Mack69}, 
and its realization via the Jordan algebra $J_2^\Cn$ of hermitean 
$2 \times 2$ matrices with the hermiticity preserving commutative 
(but non-associative) product 
\[
a\circ b := \ft12 (ab + ba)
\]
(basic properties of Jordan algebras are summarized in appendix A).
As is well known, these matrices are in one-to-one correspondence with 
four-vectors $x^\mu$ in Minkowski space via the formula 
$x \equiv x_\mu \sigma^\mu$ where $\sigma^\mu := (1, \vec{\sigma})$. 
The ``norm form'' on this algebra is just the ordinary determinant, i.e.
\[
\cN_2(x) := \det x = x_\mu x^\mu
\]
(it will be a higher order polynomial in the general case). Defining 
$\bar x := x_\mu \bar\sigma^\mu$ (where $\bar\sigma^\mu := (1,-\vec{\sigma})$)
we introduce the Jordan triple product on $J_2^\Cn$:
\[
\JTP{a}{b}{c} 
     &:=& (a\circ \bar b)\circ c + (c\circ \bar b)\circ a
           - (a \circ c) \circ \bar b \non
     &=& \ft12 ( a \bar b c + c \bar b a ) =
        \langle a,b \rangle c + \langle c , b \rangle a - 
       \langle a, c \rangle b
\]
with the standard Lorentz invariant bilinear form $\langle a, b \rangle :=
a_\mu b^\mu$. However, it is not generally true that the Jordan triple product
can be thus expressed in terms of a bilinear form.

The automorphism group of $J_2^\Cn$, which is by definition compatible with
the Jordan product, is just the rotation group $SU(2)$; the structure group,
defined as the invariance of the norm form up to a constant factor, is the
product $SL(2,\Cn) \times \mathcal{D}$, i.e. the Lorentz group and
dilatations. The conformal group associated with $J_2^\Cn$ is the group
leaving invariant the light-cone $\cN_2(x)=0$.  As is well known, the
associated Lie algebra is $su(2,2)$, and possesses a three-graded structure
\[ \mathfrak{g} = \mathfrak{g}^{-1} \oplus \mathfrak{g}^{0} 
                   \oplus\mathfrak{g}^{+1} \,, \lb{3grading}
\]
where the grade $-1$ and grade $+1$ spaces correspond to generators 
of translations $P^\mu$ and special conformal transformations $K^\mu$,
respectively, while the grade 0 subspace is spanned by the Lorentz
generators $M^{\mu\nu}$ and the dilatation generator $D$. The subspaces
$\fg^1$ and $\fg^{-1}$ can each be associated with the Jordan algebra 
$J_2^\Cn$, such that their elements are labeled by elements 
$a = a_\mu \sigma^\mu$ of $J_2^\Cn$. The precise correspondence is
\[
U_a := a_\mu P^\mu \in \fg^{-1}\qquad \text{and} \qquad 
\tilde U_a := a_\mu K^\mu \in \fg^{+1} \,.
\]
By contrast, the generators in $\mathfrak{g}^0$ are labeled by {\it two} 
elements $a,b \in J_2^\Cn$, viz.
\[
S_{a b} := a_\mu b_\nu ( M^{\mu\nu} + \eta^{\mu\nu} D) \,.
\]
The conformal group is realized non-linearly on the space of four-vectors 
$x\in J_2^\Cn$, with a M\"obius-like infinitesimal action of the special 
conformal transformations
\[
\delta  x^\mu  = 2 \langle c,x\rangle x^\mu - \langle x ,x \rangle c^\mu
\]
with parameter $c^\mu$. All variations acquire a very simple form 
when expressed in terms of above generators: we have 
\[
U_a (x) &=& a \,, \non
S_{ab} (x) &=&  \JTP{a}{b}{x}\,, \non
\tilde U_c (x) &=& -\ft12 \JTP{x}{c}{x} \,,
\]
where $\{...\}$ is the Jordan triple product introduced above.
{}From these transformations it is elementary to deduce the commutation 
relations 
\[ 
 {}[U_a, \tilde U_b]   &=& S_{a b} \, ,\non
 {}[S_{a b},U_c] &=& U_{\{abc\}} \, , \non
 {}[S_{a b} ,\tilde U_c]    &=& \tilde U_{\{bac\}} \, ,\non
 {}[S_{a b},S_{c d}] &=& S_{{\{abc\}}\, d} - S_{{\{bad\}} \, c} \,.
\]
(of course, these could have been derived directly from those of the 
conformal group). As one can also see, the Lie algebra $\fg$ admits 
an involutive automorphism $\iota$ exchanging $\fg^{-1}$ and 
$\fg^{+1}$ (hence, $\iota (K^\mu) = P^\mu$).

The above transformation rules and commutation relations exemplify the
structure that we will encounter again in section 3 of this paper: the
conformal realization of $\EE$ on $\Rn^{27}$ presented there has the same
form, except that $J_2^\Cn$ is replaced by the exceptional Jordan algebra
$J_3^{\On_S}$ over the split octonions $\On_S$.  While our form of the
nonlinear variations appears to be new, the concomitant construction of the
Lie algebra itself by means of the Jordan triple product has been known in the
literature as the Tits-Kantor-Koecher
construction~\cite{Tits62,Kant65,Koec67}, and as such generalizes
to other Jordan algebras. The generalized linear fractional
(M\"{o}bius) groups of Jordan algebras can be abstractly defined
in an analogous manner~\cite{Koec67a}, and shown to leave
invariant certain generalized $p$-angles defined by the norm form
of degree $p$ of the underlying Jordan
algebra~\cite{Kant67,Guna93}. However, to our knowledge,
explicit formulas of the type derived here have not appeared in
the literature before.

While this construction works for the exceptional Lie algebras \6[(6)], 
and \7[(7)], as well as other Lie algebras admitting a three graded 
structure, it fails for $\E$, \4[(4)], and \Gtwo[(2)], for which a
three grading does not exist. These algebras possess only a
five graded structure
\[
\fg = \fg^{-2} \oplus \fg^{-1} \oplus \fg^0 \oplus \fg^{+1} 
      \oplus \fg^{+2} \,.  \lb{5grading}
\]
Our main result, to be described in section 2, states that a
``quasiconformal'' realization is still possible on a space of
dimension ${\rm dim} (\fg^1) +1$ if the top grade spaces $\fg^{\pm
2}$ are one-dimensional. Five graded Lie algebras with this
property are closely related to the so-called Freudenthal Triple
Systems~\cite{Freu54,Meyb68}, which were originally invented to
provide alternative constructions of the exceptional Lie
groups\footnote{ The more general Kantor-Triple-Systems for which
$\fg^{\pm2}$ have more than one dimension, will not be discussed
in this paper.}.  This relation will be made very explicit in the
present paper. The novel realization of \8[(8)] which we
will arrive at, together with its natural extension to
$E_{8(-24)}$, contains various other constructions of exceptional
Lie algebras by truncation, including the conformal realizations
based on a three graded structure. For this reason, we describe
it first in section \ref{e8-realization}, and then show how the
other cases can be obtained from it.

Whereas previous attempts to construct generalized space-times mainly
focused on generalizing Minkowski space-time and its symmetries, the
physical applications that we have in mind here are of a somewhat
different nature, and inspired by recent developments in superstring
and \hbox{M-Theory}. Namely, the generalized ``space-times''
presented here could conceivably be identified with certain internal
spaces arising in supergravity and superstring theory, which are related
to the appearance of central charges in the associated superalgebras.
Central charges and their solitonic carriers have been much discussed
in the recent literature because it is hoped that they may provide a
window on \hbox{M-Theory} and its non-perturbative degrees of freedom.
More specifically, it has been argued in \cite{deWNic00} that a proper 
description of the non-perturbative \hbox{M-Theory} degrees of freedom 
might require supplementing ordinary space-time coordinates by central 
charge coordinates. Solitonic charges also play an important role 
in the microscopic description of black hole entropy: for maximally 
extended $N\!=\!8$ supergravity, the latter is conjectured
to be given by an $E_{7(7)}$ invariant formula~\cite{KalKol96,FerMal98}.
The corresponding formula for the entropy in maximally extended
supergravity in five dimensions is $E_{6(6)}$ invariant and
involves a cubic form. In~\cite{FerGun98} an invariant classification
of orbits of $E_{7(7)}$ and $E_{6(6)}$ actions on their fundamental
representations that classify BPS states in $d=4$ and $d=5$ was given.

The entropy formula in \cite{KalKol96,FerMal98} is identical to 
the equation for a vector  with vanishing norm in 57 dimensions (see 
eq.~\Eq{E8lc}), provided we use the \SLR form of the quartic $E_{7(7)}$
invariant. This suggests that the 57th component of our $\E$ realization 
should be interpreted as the entropy. However, we should stress that 
the quartic invariant can assume both positive and negative values,
cf. the simple examples given in appendix B. In order to avoid
imaginary entropy, one must therefore restrict oneself to the
positive semi-definite  values of the quartic invariant, corresponding
to the "time-like" and "light-like" orbits of $E_{7(7)}$ in the 
language of \cite{FerGun98}. With the 57th coordinate interpreted 
as the entropy and the remaining 56 coordinates as the electric 
and magnetic charges, it is natural from our point of view to
define a distance in this "entropy-charge space" between any two 
black hole solutions using our eqs.~\Eq{difference},\Eq{N4}. If two 
black hole solutions  are light-like separated in this space, they will 
remain so under the action of $\E$.\footnote{For the exceptional $N=2$
Maxwell-Einstein supergravity \cite{GuSiTo8384} defined by the
exceptional Jordan algebra the U-duality groups in five and four
dimensions are $E_{6(-26)}$ and $E_{7(-25)}$, respectively. The
quasi-conformal symmetry of the exceptional supergravity in four
dimensions is hence $E_{8(-24)}$, with the  maximal compact
subgroup  $E_7 \times SU(2)$.}. We should also point out that it is 
not entirely clear from the existing black hole literature  
whether it is the \SU or the \SLR form of the invariant that 
should be used here (the detailed relation between the two is 
worked out in appendix B). The \SU basis is relevant for the 
central charges, which appear in the superalgebra via surface 
integrals at spatial infinity and determine the structure (and length) 
of BPS multiplets. By contrast, the 28 electric and 28 magnetic 
charges carried by the solitons of $d\!=\!4,N\!=\!8$ supergravity 
transform separately under \SLR \cite{CreJul79}, and therefore the 
\SLR form of the invariant appears more appropriate in this context.

For applications to \hbox{M-Theory} it would be important to obtain the
exponentiated version of our realization. One might reasonably expect that
modular forms with respect to a fractional linear realization of the
arithmetic group $\8[(8)](\Zn)$ will have a role to play. We expect that our
results will pave the way for the explicit construction of such modular forms.
According to~\cite{Jose76} these would depend on $28\!+\!1=29$ variables, such
that the 57-dimensional Heisenberg subalgebra of $\E$ exhibited here 
would be realized in terms of 28 ``coordinates'' and 28 ``momenta''.
Consequently, the 57 dimensions in which \8[(8)] acts might alternatively 
be interpreted as a generalized Heisenberg group, in which case 
the 57th component would play the role of a variable parameter~$\hbar$. 
The action of $\8[(8)](\Zn)$ on the 57 dimensional
Heisenberg group would then constitute the invariance group of a generalized
Dirac quantization condition.  This observation is also in accord with the
fact that the term modifying the vector space addition in $\Rn^{57}$ (cf.
eq.\Eq{difference}), which is required by $\E$ invariance, is just the cocycle
induced by the standard canonical commutation relations on an
(28+28)-dimensional phase space.  

\mathversion{bold}
\section{Quasiconformal Realization of $\E$}\lb{e8-realization}
\mathversion{normal}

\mathversion{bold}
\subsection{\7[(7)] decomposition of \8[(8)] }
\mathversion{normal} 

We will start with the maximal case, the exceptional Lie group $\E$, 
and its quasiconformal realization on $\Rn^{57}$, because this 
realization contains all others by truncation. Our results are based 
on the following five graded decomposition of $\E$ with respect to
its $E_{7(7)}\times \cD$ subgroup
\[
\ba{c@{\,\,\oplus\,\,}c@{\,\,\oplus\,\,}c@{\,\,\oplus\,\,}c@{\,\,\oplus\,\,}c}
\fg^{-2} & \fg^{-1} & \fg^0 & \fg^{+1} &\fg^{+2}  \\[1ex]
\rep{1}  & \rep{56} & (\rep{133}\oplus\rep{1}) & \rep{56} & \rep{1} \ea
\lb{e8-grading}
\]
with the one-dimensional group $\cD$ consisting of dilatations. 
$\cD$ itself is part of an $SL(2,\Rn)$ group, and the above decomposition
thus corresponds to the decomposition $\rep{248}\rightarrow
(\rep{133},\rep{1})\oplus (\rep{56},\rep{2})\oplus(\rep{1},\rep{3})$
of $\E$ under its subgroup $\EE\times SL(2,\Rn)$. 

In order to write out the $\EE$ generators, it is convenient to further
decompose them w.r.t. the subgroup \SLR of \7[(7)]. In this
basis, the Lie algebra of \7[(7)] is spanned by the \SLR generators 
$G^i{}_j$, and the antisymmetric generators $G^{ijkl}$, transforming 
in the \rep{63} and \rep{70} representations of \SLR,
respectively. We also define
\[ 
G_{ijkl} &:=& \ft{1}{24}\epsd[ijklmnpq]\,G^{mnpq} \nn
\]
with \SLR indices $1\le i,j,\ldots \le 8$. 
The commutation relations are 
\[
\Com{G^i{}_j}{G^{k\vphantom{l}}{}_l} &=& 
  \gd{}_{j}^{k}\, G^i{}_{l} -\gd{}_{l}^{i}\, G^k{}_{j} \,,\non
\Com{G^i{}_j}{G^{klmn}}  &=& -4\,\gd{}_{j}^{[k}\, G^{lmn]i}_{\phantom{k}}
                             -\ft12\, \gd{}^i_j G^{klmn}_{\phantom{j}}\,,\non 
\Com{G^{ijkl}}{G^{mnpq}} &=& -\ft{1}{36}\,\eps[ijkls{[}mnp]\, 
                                 G^{q]}{}_{s} \,.\nn
\]
The fundamental \rep{56} representation of $E_{7(7)}$ is spanned by the two
antisymmetric real tensors $X^{ij}$ and $X_{ij}$ and the action of \7[(7)] is
given by\footnote{We emphasize that $X^{ij}$ and $X_{ij}$ are independent.
  This convention differs from the one used for the \SU basis in the
  appendix.}
\[
\gd X^{ij} &=& \Lambda^i{}_k X^{kj} -\Lambda^j{}_k X^{ki}  
              + \Sigma^{ijkl} X_{kl}\,, \non[1ex]
\gd X_{ij} &=& \Lambda^k{}_i X_{jk} -\Lambda^k{}_j X_{ik}  
              + \Sigma_{ijkl} X^{kl}\,,\lb{act-56} 
\]
where
\[ 
\Sigma_{ijkl} &=& \ft{1}{24} \epsd[ijklmnpq] \Sigma^{mnpq}
\]

In order to extend $\EE \times \cD$ to the full \8[(8)], we must 
enlarge $\cD$ to an $SL(2,\Rn)$ with generators $(E,F,H)$ 
in the standard Chevalley basis, together with $2\times 56$ 
further real generators $(E_{ij}, E^{ij})$ and $(F_{ij}, F^{ij})$.
Under hermitean conjugation, we have
\[
E^{ij} \;=\; F_{ij}^\dag  \,,  \qquad 
F^{ij} \;=\; -E_{ij}^\dag \,,  \quad \text{and} \quad 
E      \;=\; -F^\dag \,. \nn
\]
  
The grade $-2,-1,1$ and $2$ subspaces in the above decomposition
correspond to the subspaces $\fg^{-2}$, $\fg^{-1}$, $\fg^1$, and 
$\fg^2$ in \Eq{e8-grading}, respectively:
\[
E \oplus \{E^{ij},E_{ij}\} \oplus \{G^{ijkl},\, G^i{}_j \, ;\, H\} 
  \oplus \{F^{ij},F_{ij}\} \oplus F
\lb{e8-grading-gen}
\]
The grading may be read off from the commutators with $H$
\[
\begin{array}{rclcrcl}
\Com{H}{E}      &=& - 2\,E   \,, &\quad& \Com{H}{F}      &=& 2\,F \,,\\[1ex] 
\Com{H}{E^{ij}} &=& - E^{ij} \,, &\quad& \Com{H}{F^{ij}} &=& F^{ij}\,,\\[1ex]
\Com{H}{E_{ij}} &=& - E_{ij} \,, &\quad& \Com{H}{F_{ij}} &=& F_{ij}\,.
\end{array}\nn
\]
The new generators $(E_{ij}, E^{ij})$ and $(F_{ij}, F^{ij})$ form two
(maximal) Heisenberg subalgebras of dimension 28
\[
\begin{array}{rclcrcl}
\Com{E^{ij}}{E_{kl}} &=& 2\, \gd{}_{kl}^{ij}\, E \,,&\quad& 
\Com{F^{ij}}{F_{kl}} &=& 2\, \gd{}_{kl}^{ij}\, F \,,
\end{array}\nn
\]
and they transform under \SLR as
\[
\Com{G^i{}_j}{E^{kl}} &=&  \gd{}^k_j\,E^{il} -\gd{}^l_j\,E^{ik} 
                           -\ft14\gd{}^i_j\,E^{kl} \,,\non
\Com{G^i{}_j}{E_{kl}} &=&  \gd{}^i_k\,E_{lj} -\gd{}^i_l\,E_{kj} 
                           +\ft14\gd{}^i_j\,E_{kl} \,,\non
\Com{G^i{}_j}{F^{kl}} &=&  \gd{}^k_j\,F^{il} -\gd{}^l_j\,F^{ik}
                           -\ft14\gd{}^i_j\,F^{kl} \,,\non
\Com{G^i{}_j}{F_{kl}} &=&  \gd{}^i_k\,F_{lj} -\gd{}^i_l\,F_{kj}
                           +\ft14\gd{}^i_j\,F_{kl} \,.\nn
\]
The remaining non-vanishing commutation relations are given by
\[
[E,F]=H \nn
\]
and 
\[
\begin{array}{rclcrcl}
\!\!\!\Com{G^{ijkl}}{E_{mn}} &=& -\gd{}^{[ij}_{mn}\,E^{kl]}_{\vphantom{mn}} \,,
&& 
\Com{G^{ijkl}}{E^{mn}} &=& -\ft{1}{24}\,\eps[ijklmnpq]\,E_{pq} \,,\\[1ex]
\!\!\!\Com{G^{ijkl}}{F_{mn}} &=& -\gd{}^{[ij}_{mn}\,F^{kl]}_{\vphantom{mn}} \,,
&& 
\Com{G^{ijkl}}{F^{mn}} &=& -\ft{1}{24}\,\eps[ijklmnpq]\,F_{pq} \,,
\end{array}\nn
\]
\[
\begin{array}{rclcrcl}
\Com{E^{ij}}{F^{kl}} &=& 12\,G^{ijkl} \,,&\;& 
\Com{E_{ij}}{F_{kl}} &=& -12\,G_{ijkl} \,,\\[1ex]
\Com{E^{ij}}{F_{kl}} &=&  
 4\,\gd{}_{[k}^{[i} G^{j]}_{\phantom{[]}}{}^{\phantom{[]}}_{l]}
 -\gd{}_{kl}^{ij}\, H, &\;& 
\Com{E_{ij}}{F^{kl}} &=&  
 4\,\gd{}^{[k}_{[i} G^{l]}_{\phantom{[]}}{}^{\phantom{[]}}_{j]}
 +\gd{}_{ij}^{kl}\, H, \\[1ex]
\Com{E}{F^{ij}} &=& -E^{ij} \,, &\;& 
\Com{E}{F_{ij}} &=& -E_{ij} \,,\\[1ex] 
\Com{F}{E^{ij}} &=& F^{ij} \,,&\;& 
\Com{F}{E_{ij}} &=& F_{ij} \,.
\end{array}\nn
\]
To see that we are really dealing with the maximally split form of $\E$, let
us count the number of compact generators: The antisymmetric part
$(G^i{}_j-G^j{}_i)$ of $G^i{}_j$ and $(G^{ijkl}-G_{ijkl})$ correspond to the
63 generators of of the maximal compact supalgebra $SU(8)$ of
\7[(7)]~\cite{CreJul79}. The remaining compact generators are the $28+28+1$
anti-hermitean generators $(E_{ij}+F^{ij})$, $(E^{ij}-F_{ij})$, and $(E+F)$
giving a total of 120 generators which close into the maximal compact subgroup
$\SO[16]\supset\SU$ of $\E$.

An important role is played by the symplectic invariant of two \rep{56}
representations. It is given by
\[ 
 \Sympl{X}{Y} &:=& X^{ij}Y_{ij}-X_{ij}Y^{ij}\,.
 \lb{sympl56-def}
\]
The second structure which we need to introduce is the triple product. This is
a trilinear map $\rep{56}\times\rep{56}\times\rep{56}\longrightarrow\rep{56}$,
which associates to three elements $X$, $Y$ and $Z$ another element
transforming in the $\rep{56}$ representation, denoted by $\FTP{X}{Y}{Z}$, and
defined by
\[ \FTP[ij]{X}{Y}{Z} &:=& 
  -8\,X\oversym{{}^{ik}Y_{kl}Z^{lj}}
  -8\,Y\oversym{{}^{ik}X_{kl}Z^{lj}}
  -8\,Y\oversym{{}^{ik}Z_{kl}X^{lj}}\non[1ex]
&&
  -2\,Y^{ij}X^{kl}Z_{kl}
  -2\,X^{ij}Y^{kl}Z_{kl}
  -2\,Z^{ij}Y^{kl}X_{kl} \non[1ex]
&&
  +\ft{1}{2}\,\eps[ijklmnpq]X_{kl}Y_{mn}Z_{pq} \non[1ex]
\FTP{X}{Y}{Z}_{ij} &:=&
   8\,X\undersym{{}_{ik}Y^{kl}Z_{lj}}
  +8\,Y\undersym{{}_{ik}X^{kl}Z_{lj}}
  +8\,Y\undersym{{}_{ik}Z^{kl}X_{lj}}\non[1ex]
&&
  +2\,Y_{ij}Z^{kl}X_{kl}
  +2\,X_{ij}Z^{kl}Y_{kl}
  +2\,Z_{ij}X^{kl}Y_{kl} \non[1ex]
&&
  -\ft{1}{2}\,\epsd[ijklmnpq]X^{kl}Y^{mn}Z^{pq}\,.
  \lb{ftp56-def}
\]
A somewhat tedious calculation\footnote{Which relies heavily on the 
Schouten identity $\varepsilon_{[ijklmnpq} X_{r]s} = 0.$} shows that 
this triple product obeys the relations
\[
\FTP{X}{Y}{Z} &=& \FTP{Y}{X}{Z} +2\,\Sympl{X}{Y}Z \,,\non[1ex]
\FTP{X}{Y}{Z} &=& \FTP{Z}{Y}{X} -2\,\Sympl{X}{Z}Y \,,\non[1ex]
\Sympl{\FTP{X}{Y}{Z}}{W} &=& \Sympl{\FTP{X}{W}{Z}}{Y} 
                                -2\,\Sympl{X}{Z}\Sympl{Y}{W} \,,\non[1ex]
\FTP{X}{Y}{\FTP{V}{W}{Z}} &=& \FTP{V}{W}{\FTP{X}{Y}{Z}}
                             +\FTP{\FTP{X}{Y}{V}}{W}{Z} \non
                          && {}+\FTP{V}{\FTP{Y}{X}{W}}{Z} \,.\lb{ftp56-rel}
\]
We note that the triple product \Eq{ftp56-def} could be modified by 
terms involving the symplectic invariant, such as $\langle X,Y \rangle Z$;
the above choice has been made in order to obtain agreement with
the formulas of~\cite{Faul71}.

While there is no (symmetric) quadratic invariant of $\EE$ in the $\rep{56}$
representation, a real quartic invariant $\cI_4$ can be constructed
by means of the above triple product and the bilinear form; it reads 
\[
\cI_4(X^{ij},X_{ij}) 
&:=&\ft{1}{48}\,\Sympl{\FTP{X}{X}{X}}{X} \non[1ex]
&\=& X^{ij}X_{jk}X^{kl}X_{li} 
    -\ft14 X^{ij}X_{ij}X^{kl}X_{kl} \non
&&
 +\ft{1}{96}\,\eps[ijklmnpq]X_{ij}X_{kl}X_{mn}X_{pq} \non
&&
 +\ft{1}{96}\,\epsd[ijklmnpq]X^{ij}X^{kl}X^{mn}X^{pq} \lb{e7-inv} \,.
\lb{e7-invariant}
\]

\mathversion{bold}
\subsection{Quasiconformal nonlinear realization of $E_{8(8)}$}
\mathversion{normal} 

We will now exhibit a nonlinear realization of $\E$ on the $57$-dimensional
real vector space with coordinates
\[\cX:=(X^{ij},X_{ij},x) \,, \nn\] 
where $x$ is also real. While $x$ is a $\EE$ singlet, the remaining 56
variables transform linearly under $\EE$. Thus $\cX$ forms the
$\rep{56}\oplus\rep{1}$ representation of \7[(7)]. In writing the
transformation rules we will omit the transformation parameters in
order not to make the formulas (and notation) too cumbersome. To recover the
infinitesimal variations, one must simply contract the formulas with the
appropriate transformation parameters. The $\EE$ subalgebra acts linearly by
\[
\begin{array}{rclrcl}
G^i{}_j(X^{kl}) &=& 2\,\gd\oversym{{}^k_j X^{il}} -\ft{1}{4}\gd^i_j X^{kl}\,,
&
G^{ijkl}(X^{mn})&=& \ft{1}{24}\eps[ijklmnpq] X_{pq} \,,
\\[1ex]
G^i{}_j(X_{kl}) &=& -2\,\gd\undersym{{}^i_k X_{jl}}+\ft{1}{4}\gd^i_j X_{kl}\,,
& 
G^{ijkl}(X_{mn})&=& \gd^{[ij}_{mn} X^{kl]}_{\phantom{m}} \,,
\\[1ex]
G^i{}_j(x)    &=& 0 \,, 
& 
G^{ijkl}(x)   &=& 0 \,, 
\end{array}\lb{e8-G}
\]
$H$ generates scale transformations
\[
\begin{array}{rclcrclcrcl}
H (X^{ij}) &=&  X^{ij} \,,
&\quad&
H (X_{ij}) &=&  X_{ij} \,,
&\quad& 
H (x)    &=& 2\, x \,,
\end{array}\lb{e8-H}
\]
and the $E$ generators act as translations; we have
\[ 
\begin{array}{rclcrclcrcl}
E(X^{ij}) &=& 0 \,, &\quad&
E(X_{ij}) &=& 0 \,, &\quad& 
E(x) &=& 1 \,.
\end{array}\lb{e8-E}
\]
and 
\[ 
\begin{array}{rclcrclcrcl}
E^{ij}(X^{kl}) &=& 0 \,, &\quad&
E^{ij}(X_{kl}) &=& \gd{}^{ij}_{kl}  \,, &\quad&
E^{ij}(x)    &=& - X^{ij} \,, \\[1ex]
E_{ij}(X^{kl}) &=& \gd{}^{kl}_{ij} \,, &\quad&
E_{ij}(X_{kl}) &=& 0 \,, &\quad&
E_{ij}(x)    &=& X_{ij}\,.
\end{array}\lb{e8-Eij}
\]
By contrast, the $F$ generators are realized nonlinearly:
\[
F(X^{ij}) &=& -\ft16\FTP[ij]{X}{X}{X} + X^{ij}\,x \non[1ex]
&\=&
 4 X\oversym{{}^{ik} X_{kl} X^{lj}}
 +X^{ij} X^{kl} X_{kl} \non
&&
 -\ft{1}{12}\eps[ijklmnpq] X_{kl} X_{mn} X_{pq}
 +X^{ij}\,x \,, \non[1ex]
F(X_{ij}) &=& -\ft16\FTP{X}{X}{X}_{ij} + X_{ij}\,x \non[1ex]
&\=&
 -4 X\undersym{{}_{ik} X^{kl} X_{lj}}
 -X_{ij} X^{kl} X_{kl} \non
&&
 +\ft{1}{12}\epsd[ijklmnpq] X^{kl} X^{mn} X^{pq}
 +X_{ij}\,x \,, \non[1ex]
F(x) &=& 4\,\cI_4( X^{ij}, X_{ij})+x^2 \non[1ex]
&\=&
  4\, X^{ij} X_{jk} X^{kl} X_{li} 
 - X^{ij} X_{ij} X^{kl} X_{kl} \non
&&
 +\ft{1}{24}\,\eps[ijklmnpq] X_{ij} X_{kl} X_{mn} X_{pq} \non
&&
 +\ft{1}{24}\,\epsd[ijklmnpq] X^{ij} X^{kl} X^{mn} X^{pq}
 +x^2 \,.\lb{e8-F}
\]
Observe that the form of the r.h.s. is dictated by the requirement
of $\EE$ covariance: $(F(X^{ij}),F(X_{ij}))$ and $F(x)$ must 
still transform as the $\rep{56}$ and $\rep{1}$ of $\EE$, respectively.
The action of the remaining generators is likewise $\EE$ covariant: 
\[
F^{ij}(X^{kl}) &=&
 -4\, X\oversym{{}^{i[k} X^{l]j}}
 +\ft{1}{4}\,\eps[ijklmnpq] X_{mn} X_{pq} \,, \non[1ex]
F^{ij}(X_{kl}) &=&
 +8\,\gd\undersym{{}^{[i}_k X^{j]m}_{\phantom{m]n}} X_{ml}}
 +\gd{}^{ij}_{kl}\, X^{mn} X_{mn}   +2\, X^{ij} X_{kl}
 -\gd{}^{ij}_{kl}\,x \,, \non[1ex]
F_{ij}(X^{kl}) &=&
 -8\,\gd\oversym{{}_{[i}^k X_{j]m}^{\phantom{m}} X^{ml}}
 +\gd{}^{kl}_{ij}\, X^{mn} X_{mn}   -2\, X_{ij} X^{kl}
 -\gd{}^{kl}_{ij}\,x \,, \non[1ex]
F_{ij}(X_{kl}) &=&
 4\, X\undersym{{}_{ki} X_{jl}}
 -\ft{1}{4}\,\epsd[ijklmnpq] X^{mn} X^{pq} \,, \non[1ex]
F^{ij}(x) &=& 
 4\, X\oversym{{}^{ik} X_{kl} X^{lj}} + X^{ij} X^{kl} X_{kl} \non 
&&-\ft{1}{12}\,\eps[ijklmnpq]  X_{kl} X_{mn} X_{pq}
 + X^{ij}\,x \,, \non[1ex]
F_{ij}(x) &=& 
 4\, X\undersym{{}_{ik} X^{kl} X_{lj}} + X_{ij} X^{kl} X_{kl} \non 
&&-\ft{1}{12}\,\epsd[ijklmnpq]  X^{kl} X^{mn} X^{pq}
 - X_{ij}\,x  \,. \lb{e8-Fij}
\]
Although $\EE$ covariance considerably constrains the expressions
that can appear on the r.h.s., it does not fix them uniquely:
as for the triple product \Eq{ftp56-def} one could add further
terms involving the symplectic invariant. However, all ambiguities
are removed by imposing closure of the algebra, and we have checked
by explicit computation that the above variations do close into
the full $\E$ algebra in the basis given in the previous section.
This is the crucial consistency check.

The term ``quasiconformal realization'' is motivated by the existence
of a norm form that is left invariant up to a (possibly coordinate
dependent) factor under all transformations. To write it down we
must first define a nonlinear ``difference'' between two points
$\cX\= (X^{ij},X_{ij}\, ;\,x)$ and $\cY\= (Y^{ij},Y_{ij}\, ;\,y)$;
curiously, the standard difference is {\it not} invariant under the 
translations $(E^{ij},E_{ij})$. Rather, we must choose
\[
\gd(\cX,\,\cY) \;:=\; (X^{ij}-Y^{ij},X_{ij}-Y_{ij}\;;\;x-y+\Sympl{X}{Y})\,.
\lb{difference}
\] 
This difference still obeys $\delta(\cX,\cY)\!=\!-\delta(\cY,\cX)$ and
thus $\delta(\cX,\cX)\!=\!0$, and is now invariant under $(E^{ij},E_{ij})$ 
as well as $E$; however, it is no longer additive. In fact, with
the sum of two vectors being defined as $\gd(\cX,-\cY)$, the extra
term involving $\langle X,Y \rangle$ can be interpreted as the 
cocycle induced by the standard canonical commutation relations.

The relevant invariant is a linear combination of $x^2$ and the
quartic $\EE$ invariant $\cI_4$, viz.
\[
 \cN_4(\cX)\;\=\;\cN_4( X^{ij}, X_{ij}; x) \;:=\; \cI_4( X) - x^2\,, \lb{N4}
\]
In order to ensure invariance under the translation generators,
we consider the expression $\cN_4(\gd(\cX,\cY))$, which is 
manifestly invariant under the linearly realized subgroup $\EE$.
Remarkably, it also transforms into itself up to an overall factor 
under the action of the nonlinearly realized generators. 
More specifically, we find
\[
F\Big(\cN_4(\gd(\cX,\cY))\Big) &=& 2\,(x+y)\,\cN_4(\gd(\cX,\cY)) \non
F^{ij}\Big(\cN_4(\gd(\cX,\cY))\Big) &=& 
  2\,(X^{ij}+Y^{ij})\,\cN_4(\gd(\cX,\cY)) \non
H\Big(\cN_4(\gd(\cX,\cY))\Big) &=& 4\,\cN_4(\gd(\cX,\cY)) \nn
\]
Therefore, for every $\cY\in \Rn^{57}$ the ``light cone'' with base 
point $\cY$, defined by the set of $\cX\!\in\Rn^{57}$ obeying
\[
\cN_4(\gd(\cX,\cY)) = 0\,,  \lb{E8lc}
\]
is preserved by the full $\E$ group, and in this sense, $\cN_4$ is a
``conformal invariant'' of $\E$. We note that the light cones defined by the
above equation are not only curved hypersufaces in $\Rn^{57}$, but get
deformed as one varies the base point $\cY$.  As we will show in appendix B,
the quartic invariant $\cI_4$ can take both positive and negative values,
but in the latter case eq.~\Eq{E8lc} does not have real solutions. However, we
can remedy this problem by extending the representation space to $\Cn^{57}$
and using the same formulas to get a realization of the complexified Lie
algebra $\8(\Cn)$ on $\Cn^{57}$. 

The existence of a fourth order conformal invariant of $\E$ is 
noteworthy in view of the fact that no irreducible fourth 
order invariant exists for the linearly realized $\E$ group (the
next invariant after the quadratic Casimir being of order eight).

\mathversion{bold}
\subsection{Relation with Freudenthal Triple Systems}
\mathversion{normal}

We will now rewrite the nonlinear transformation rules in another
form in order to establish contact with mathematical literature. 
Both the bilinear form \Eq{sympl56-def} and the triple product 
\Eq{ftp56-def} already appear in~\cite{Faul71}, albeit in a very
different guise. That work starts from $2\times 2$ ``matrices'' of 
the form
\[
A = \left( 
\begin{array}{cc}
  \alpha_1 & x_1 \\ x_2 & \alpha_2
\end{array}   \right)\,,
\]
where $\alpha_1,\alpha_2$ are real numbers and $x_1,x_2$ are elements of a
simple Jordan algebra $J$ of degree three. There are only four simple  
Jordan algebras $J$ of this type, namely the $3 \times 3$ hermitean 
matrices over the four division algebras, $\Rn$, $\Cn$, $\Hn$ and $\On$. 
The associated matrices are then related to non-compact forms of
the exceptional Lie algebras \4, \6, \7, and \8, respectively.
For simplicity, let us concentrate on the maximal case $J_3^{\On_S}$,
when the matrix $A$ carries 1+1+27+27 = 56 degrees of freedom. 
This counting suggests an obvious relation with the $\rep{56}$ 
of $\EE$ and its decomposition under $\EEE$, but more work is required 
to make the connection precise. To this aim, \cite{Faul71} defines
a symplectic invariant $\langle A , B \rangle$, and a trilinear 
product mapping three such matrices $A,B$ and $C$ to another one, 
denoted by $\FTP{A}{B}{C}$. This triple system differs from a
Jordan triple system in that it is not derivable from a binary
product. The formulas for the triple product in terms of the 
matrices $A,B$ and $C$ given in~\cite{Faul71} are somewhat cumbersome, 
lacking manifest $\EE$ covariance. For this reason, instead of 
directly verifying that our prescription \Eq{ftp56-def} and the 
one of \cite{Faul71} coincide, we have checked that they
satisfy identical relations: a quick glance shows that the 
relations (T1)--(T4)~\cite{Faul71} are indeed the same as our
relations \Eq{ftp56-rel}, which are manifestly $\EE$ covariant. 

To rewrite the transformation formulas we introduce Lie algebra generators
$U_A$ and $\tilde U_A$ labeled by the above matrices, as well as
generators $S_{AB}$ labeled by a pair of such matrices. For the grade 
$\pm 2$ subspaces we would in general need another set of generators 
$K_{AB}$ and $\tilde K_{AB}$ labeled by two matrices, but since these 
subspaces are one-dimensional in the present case, we have
only two more generators $K_a$ and $\tilde K_a$ labelled by one 
real number $a$. In the same vein, we reinterpret the 57 coordinates 
$\cX$ as a pair $( X,x)$, where $ X$ is a $2\times 2$ matrix of
the type defined above. The variations then take the simple form
\[
K_a (X)     &=& 0             \,, \non[1ex]
K_a (x)     &=& 2\, a             \,, \non[1ex]
U_A (X)     &=& A             \,, \non[1ex]
U_A (x)     &=& \Sympl{A}{X}  \,, \non[1ex]
S_{AB} (X)     &=& \FTP{A}{B}{X} \,, \non[1ex] 
S_{AB} (x)     &=& 2\Sympl{A}{B} x   \,, \non[1ex]
\tilde U_A (X) &=& \2\FTP{X}{A}{X} - A x   \,, \non[1ex]
\tilde U_A (x) &=& -\ft16\Sympl{\FTP{X}{X}{X}}{A}
                   +\Sympl{X}{A}x         \,, \non[1ex]
\tilde K_a (X) &=& -\ft16\,a\FTP{X}{X}{X} +a Xx \,, \non[1ex]
\tilde K_a (x) &=&  \ft16\,a \Sympl{\FTP{X}{X}{X}}{X} +2\,a x^2 \,,
\]

{}From these formulas it is straightforward to determine the commutation 
relations of the transformations. To expose the connection with the more
general Kantor triple systems we write
\[
K_{AB} \equiv K_{\langle A , B\rangle}  \lb{KAB}
\]
in the formulas below. The consistency of this specialization is
ensured by the relations \Eq{ftp56-rel}. By explicit computation one finds
\[
 {}[U_A,\tilde{U}_B]         &=& S_{AB}             \,, \non[1ex]
 {}[U_A,U_B]                 &=& -K_{AB}         \,, \non[1ex]
 {}[\tilde{U}_A,\tilde{U}_B] &=& -\tilde{K}_{AB} \,, \non[1ex]
 {}[S_{AB},U_C]            &=& -U_{\FTP{A}{B}{C}}          \,, \non[1ex]
 {}[S_{AB},\tilde{U}_C]    &=& - \tilde{U}_{\FTP{B}{A}{C}} \,, \non[1ex]
 {}[K_{AB},\tilde{U}_C] &=& U_{\FTP{A}{C}B} - U_{\FTP{B}{C}A} \,, \non[1ex]
 {}[\tilde{K}_{AB},U_C] &=& \tilde{U}_{\FTP{B}{C}{A}}
                           -\tilde{U}_{\FTP{A}{C}{B}} \,, \non[1ex]
 {}[S_{AB},S_{CD}]      &=& -S_{\FTP{A}{B}{C}D} 
                            -S_{C\FTP{B}{A}{D}} \,, \non[1ex]
 {}[S_{AB},K_{CD}]         &=&  K_{A\FTP{C}{B}{D}} 
                              - K_{A\FTP{D}{B}{C}} \,,\non[1ex]
 {}[S_{AB},\tilde{K}_{CD}] &=&  \tilde{K}_{\FTP{D}{A}{C}B}
                              - \tilde{K}_{\FTP{C}{A}{D}B}\,, \non[1ex]
 {}[K_{AB},\tilde{K}_{CD}] &=&   S_{\FTP{B}{C}{A}D}
                                -S_{\FTP{A}{C}{B}D}
                                -S_{\FTP{B}{D}{A}C}
                                +S_{\FTP{A}{D}{B}C} \,.
\]
For general $K_{AB}$, these are the defining commutation relations
of a Kantor triple system, and, with the further specification
\Eq{KAB}, those of a Freudenthal triple system (FTS).  Freudenthal
introduced these triple systems in his study of the metasymplectic
geometries associated with exceptional groups~\cite{Freu85}; these
geometries were further studied
in~\cite{AllFau84,Faul71,Meyb68,KanSko82}\footnote{FTS's have also
been used in \cite{BinGun97} to give a classification and a unified
realization of non-linear quasi-superconformal algebras and in the
realizations of nonlinear $N=4$ superconformal algebras in two
dimensions \cite{GunHyu9293}.}. A classification of FTS's may be found
in~\cite{KanSko82}, where it is also shown that there is a
one-to-one correspondence between simple Lie algebras and simple
FTS's with a non-degenerate bilinear form. Hence there is a
quasiconformal realization of every Lie group acting on a
generalized lightcone.

\mathversion{bold}
\section{Truncations of $\E$}
\mathversion{normal}

For the lower rank exceptional groups contained in $\E$, we can derive similar
conformal or quasiconformal realizations by truncation. In this section, we
will first give the list of quasiconformal realizations contained in $\E$. In
the second part of this section, we consider truncations to a three graded
structure, which will yield conformal realizations. In particular, we will
work out the conformal realization of $\EE$ on a space of 27 dimensions as an
example, which is again the maximal example of its kind.

\subsection{More quasiconformal realizations}

All simple Lie algebras (except for $SU(2)$) can be given a five graded
structure \Eq{5grading} with respect to some subalgebra of maximal rank and
associate a triple system with the grade $+1$ subspace~\cite{Kant73,BarGun79}.
Conversely, one can construct every simple Lie algebra over the corresponding
triple system.

The realization of \8 over the FTS defined by the exceptional Jordan algebra
can be truncated to the realizations of \7, \6, and \4 by restricting oneself
to subalgebras defined by quaternionic, complex, and real Hermitean $3\times
3$ matrices.  Analogously the non-linear realization of \8[(8)] given in the
previous section can be truncated to non-linear realizations of \7[(7)],
\6[(6)], and \4[(4)]. These truncations preserve the five grading. More
specifically we find that the Lie algebra of \7[(7)] has a five grading of the
form:
\[
  \7[(7)]  = \rep{\overline{1}} \oplus \rep{\overline{32}} 
             \oplus (SO(6,6)\oplus\cD)  \oplus \rep{32} \oplus \rep{1}
\]
Hence this truncation leads to a nonlinear realization of \7[(7)] on a
\rep{33} dimensional space. Note that this is not a minimal realization of
\7[(7)]. Further truncation to the \6[(6)] subgroup preserving the five
grading leads to:
\[ 
  \6[(6)]  = \rep{\overline{1}} \oplus \rep{\overline{20}} 
             \oplus (SL(6,\Rn)\oplus\cD) \oplus \rep{20} \oplus \rep{1}
\]
This yields a nonlinear realization of \6[(6)] on a \rep{21} dimensional
space, which again is not the minimal realization. Further reduction to
\4[(4)] preserving the five grading
\[
  \4[(4)]  = \rep{\overline{1}} \oplus \rep{\overline{14}} 
             \oplus (Sp(6,\Rn)\oplus\cD) \oplus \rep{14} \oplus \rep{1}
\]
leads to a minimal realization of \4[(4)] on a fifteen dimensional space. One
can further truncate \4 to a subalgebra \Gtwo[(2)] while preserving the five
grading
\[
  \Gtwo[(2)]  = \rep{\overline{1}} \oplus \rep{\overline{4}} 
                \oplus (SL(2,\Rn)\oplus\cD) \oplus \rep{4} \oplus \rep{1} \,,
\]
which then yields a nonlinear realization over a five dimensional space. One
can go even futher and truncate \Gtwo to its subalgebra $SL(3,\Rn)$
\[
  SL(3,\Rn)  = \rep{\overline{1}} \oplus \rep{\overline{2}}
               \oplus (SO(1,1)\oplus\cD) \oplus \rep{2} \oplus \rep{1} \,,
\]
which is the smallest simple Lie algebra admitting a five grading. We should
perhaps stress that the nonlinear realizations given above are minimal for
\Gtwo[(2)], \4[(4)], and \8[(8)] which are the only simple Lie algebras that do
not admit a three grading and hence do not have unitary representations of the
lowest weight type.

The above nonlinear realizations of the exceptional Lie algebras can also be
truncated to subalgebras with a three graded structure, in which case our
nonlinear realization reduces to the standard nonlinear realization over a
JTS. This truncation we will describe in section \ref{e7-real-sec} in more
detail.

With respect to \6[(6)] the quasiconformal realization of \8[(8)]
\Eq{e8-grading} decomposes as follows:
\smallskip
\begin{center}
%
%
\begin{picture}(0,0)%
\includegraphics{e7e8emb.pst}%
\end{picture}%
\setlength{\unitlength}{2486sp}%
\begingroup\makeatletter\ifx\SetFigFont\undefined%
\gdef\SetFigFont#1#2#3#4#5{%
  \reset@font\fontsize{#1}{#2pt}%
  \fontfamily{#3}\fontseries{#4}\fontshape{#5}%
  \selectfont}%
\fi\endgroup%
\begin{picture}(8100,4215)(451,-4246)
\put(451,-151){\makebox(0,0)[lb]{\smash{\SetFigFont{10}{13.2}{\rmdefault}{\mddefault}{\updefault}
$\rep{1}$
}}}
\put(1351,-151){\makebox(0,0)[lb]{\smash{\SetFigFont{10}{13.2}{\rmdefault}{\mddefault}{\updefault}
$\oplus$
}}}
\put(2251,-151){\makebox(0,0)[lb]{\smash{\SetFigFont{10}{13.2}{\rmdefault}{\mddefault}{\updefault}
$\rep{56}$
}}}
\put(3151,-151){\makebox(0,0)[lb]{\smash{\SetFigFont{10}{13.2}{\rmdefault}{\mddefault}{\updefault}
$\oplus$
}}}
\put(4051,-151){\makebox(0,0)[lb]{\smash{\SetFigFont{10}{13.2}{\rmdefault}{\mddefault}{\updefault}
$(\rep{133}\oplus\rep{1})$
}}}
\put(8551,-151){\makebox(0,0)[lb]{\smash{\SetFigFont{10}{13.2}{\rmdefault}{\mddefault}{\updefault}
$\rep{1}$
}}}
\put(7651,-151){\makebox(0,0)[lb]{\smash{\SetFigFont{10}{13.2}{\rmdefault}{\mddefault}{\updefault}
$\oplus$
}}}
\put(6751,-151){\makebox(0,0)[lb]{\smash{\SetFigFont{10}{13.2}{\rmdefault}{\mddefault}{\updefault}
$\rep{56}$
}}}
\put(5851,-151){\makebox(0,0)[lb]{\smash{\SetFigFont{10}{13.2}{\rmdefault}{\mddefault}{\updefault}
$\oplus$
}}}
\put(2341,-1456){\makebox(0,0)[lb]{\smash{\SetFigFont{10}{13.2}{\rmdefault}{\mddefault}{\updefault}
$\oplus$
}}}
\put(2341,-1051){\makebox(0,0)[lb]{\smash{\SetFigFont{10}{13.2}{\rmdefault}{\mddefault}{\updefault}
$\rep{1}$
}}}
\put(2251,-1951){\makebox(0,0)[lb]{\smash{\SetFigFont{10}{13.2}{\rmdefault}{\mddefault}{\updefault}
$\rep{27}$
}}}
\put(2251,-2851){\makebox(0,0)[lb]{\smash{\SetFigFont{10}{13.2}{\rmdefault}{\mddefault}{\updefault}
$\rep{\overline{27}}$
}}}
\put(2341,-2401){\makebox(0,0)[lb]{\smash{\SetFigFont{10}{13.2}{\rmdefault}{\mddefault}{\updefault}
$\oplus$
}}}
\put(2341,-3301){\makebox(0,0)[lb]{\smash{\SetFigFont{10}{13.2}{\rmdefault}{\mddefault}{\updefault}
$\oplus$
}}}
\put(2341,-3751){\makebox(0,0)[lb]{\smash{\SetFigFont{10}{13.2}{\rmdefault}{\mddefault}{\updefault}
$\rep{1}$
}}}
\put(6841,-3751){\makebox(0,0)[lb]{\smash{\SetFigFont{10}{13.2}{\rmdefault}{\mddefault}{\updefault}
$\rep{1}$
}}}
\put(6841,-3301){\makebox(0,0)[lb]{\smash{\SetFigFont{10}{13.2}{\rmdefault}{\mddefault}{\updefault}
$\oplus$
}}}
\put(6841,-2401){\makebox(0,0)[lb]{\smash{\SetFigFont{10}{13.2}{\rmdefault}{\mddefault}{\updefault}
$\oplus$
}}}
\put(6841,-1501){\makebox(0,0)[lb]{\smash{\SetFigFont{10}{13.2}{\rmdefault}{\mddefault}{\updefault}
$\oplus$
}}}
\put(6841,-1051){\makebox(0,0)[lb]{\smash{\SetFigFont{10}{13.2}{\rmdefault}{\mddefault}{\updefault}
$\rep{1}$
}}}
\put(451,-2401){\makebox(0,0)[lb]{\smash{\SetFigFont{10}{13.2}{\rmdefault}{\mddefault}{\updefault}
$\rep{1}$
}}}
\put(8551,-2401){\makebox(0,0)[lb]{\smash{\SetFigFont{10}{13.2}{\rmdefault}{\mddefault}{\updefault}
$\rep{1}$
}}}
\put(6751,-1951){\makebox(0,0)[lb]{\smash{\SetFigFont{10}{13.2}{\rmdefault}{\mddefault}{\updefault}
$\rep{27}$
}}}
\put(6751,-2851){\makebox(0,0)[lb]{\smash{\SetFigFont{10}{13.2}{\rmdefault}{\mddefault}{\updefault}
$\rep{\overline{27}}$
}}}
\put(4681,-601){\makebox(0,0)[lb]{\smash{\SetFigFont{10}{13.2}{\rmdefault}{\mddefault}{\updefault}
$\rep{1}$
}}}
\put(4681,-1051){\makebox(0,0)[lb]{\smash{\SetFigFont{10}{13.2}{\rmdefault}{\mddefault}{\updefault}
$\oplus$
}}}
\put(4591,-1501){\makebox(0,0)[lb]{\smash{\SetFigFont{10}{13.2}{\rmdefault}{\mddefault}{\updefault}
$\rep{27}$
}}}
\put(4681,-1951){\makebox(0,0)[lb]{\smash{\SetFigFont{10}{13.2}{\rmdefault}{\mddefault}{\updefault}
$\oplus$
}}}
\put(4591,-2401){\makebox(0,0)[lb]{\smash{\SetFigFont{10}{13.2}{\rmdefault}{\mddefault}{\updefault}
$\rep{78}$
}}}
\put(4681,-2851){\makebox(0,0)[lb]{\smash{\SetFigFont{10}{13.2}{\rmdefault}{\mddefault}{\updefault}
$\oplus$
}}}
\put(4591,-3301){\makebox(0,0)[lb]{\smash{\SetFigFont{10}{13.2}{\rmdefault}{\mddefault}{\updefault}
$\rep{\overline{27}}$
}}}
\put(4681,-3751){\makebox(0,0)[lb]{\smash{\SetFigFont{10}{13.2}{\rmdefault}{\mddefault}{\updefault}
$\oplus$
}}}
\put(4681,-4201){\makebox(0,0)[lb]{\smash{\SetFigFont{10}{13.2}{\rmdefault}{\mddefault}{\updefault}
$\rep{1}$
}}}
\end{picture}

\end{center}
The numbers in the first line are the dimensions of $\EE$, whereas the
remaining numbers correspond to representations of $\USp$ which is the maximal
compact subgroup of $\EEE$. The \rep{27} of grade $-1$ subspace and the
\rep{\overline{27}} of grade $+1$ subspace close into the $\6[(6)]\oplus\cD$
subalgebra of grade zero subspace and generate the Lie algebra of
$\EE$. Similarly \rep{\overline{27}} of grade $-1$ subspace together with
the \rep{27} of grade $+1$ subspace form another $\EE$ subalgebra of
$\E$. Hence we have four different $\EE$ subalgebras of $\E$:
%
%
\begin{enumerate}
\item[i)] \7[(7)] subalgebra of grade zero subspace which is realized
  linearly.
\item[ii)] \7[(7)] subalgebra preserving the 5-grading, which is realized nonlinearly over a 33 dimensional space  
\item[iii)] \7[(7)] subalgebra that acts on the \rep{27} dimensional subspace
  as the generalized conformal generators.
\item[iv)] \7[(7)] subalgebra that acts on the \rep{\overline{27}} dimensional
  subspace as the generalized conformal generators.
\end{enumerate}

Similarly for \7[(7)] under the $SL(6,\Rn)$ subalgebra of the grade zero
subspace the \rep{32} dimensional grade $+1$ subspace decomposes as
\[ 
  \rep{32} = \rep{1} + \rep{\overline{15}} + \rep{15} + \rep{1} \,. \nn
\] 

The \rep{15} from grade $+1\,(-1)$ subspace together with \rep{\overline{15}}
(\rep{15}) of grade $-1\,(+1)$ subspace generate a nonlinearly realized
$SO(6,6)$ subalgebra that acts as the generalized conformal algebra on the
\rep{15} (\rep{\overline{15}}) dimensional subspace.

For \6[(6)], \4[(4)], \Gtwo[(2)], and $SL(3,\Rn)$ the analogous truncations
lead to nonlinear conformal subalgebras $SL(6,\Rn)$, $Sp(6,\Rn)$, $SO(2,2)$,
and $SL(2,\Rn)$, respectively.

\mathversion{bold}
\subsection{Conformal Realization of $\EE$} \lb{e7-real-sec}
\mathversion{normal} As a special truncation the quasiconformal realization of
$\E$ contains a conformal realization of $\EE$ on a space of 27 dimensions, on
which the $\EEE$ subgroup of $\EE$ acts linearly. The main difference is that
the construction is now based on a three-graded decomposition \Eq{3grading}
of $\EE$ rather than \Eq{5grading} -- hence the realization is ``conformal''
rather than ``quasiconformal''. The relevant decomposition can be
directly read off from the figure: we simply truncate to an $\EE$
subalgebra in such a way  that the grade $\pm 2$ subspace can no 
longer be reached by commutation. This requirement is met only by
the two truncations corresponding to the diagonal lines in the figure; 
adding a singlet we arrive at the desired three graded decomposition of $\EE$
\[
\rep{133} = \rep{27}\oplus(\rep{78}\oplus\rep{1})\oplus\rep{\overline{27}}
\]
under its $\EEE \times \cD$ subgroup.

The Lie algebra \6[(6)] has \USp as its maximal compact subalgebra.  It is
spanned by a symmetric tensor $\Gt^{ij}$ in the adjoint representation
\rep{36} of \USp and a fully antisymmetric symplectic traceless tensor
$\Gt^{ijkl}$ transforming under the \rep{42} of \USp ; indices $1\le
i,j,\ldots \le 8$ are now \USp indices and all tensors with a tilde transform
under \USp rather then \SLR.  $\Gt^{ijkl}$ is traceless with respect to the
real symplectic metric $\gOd[ij]\!=\!-\gOd[ji]\!=\!-\gO[ij]$ (thus $\gOd[ik]
\gO[kj]\!=\!\delta_i^j$).  The symplectic metric also serves to pull up and
down indices, with the convention that this is always to be done from the left.

The remaining part of $\EE$ is spanned by an extra dilatation generator $\Ht$,
translation generators $\Et^{ij}$ and the nonlinearly realized generators
$\Ft^{ij}$, transforming as $\rep{27}$ and $\rep{\overline{27}}$,
respectively. Unlike for $\E$, there is no need here to distinguish the
generators by the position of their indices, since the corresponding
generators are linearly related by means of the symplectic metric.

The fundamental \rep{27} of \6[(6)] (on which we are going to realize
a nonlinear action of \7[(7)]) is given by the traceless antisymmetric 
tensor $\Zt^{ij}$ transforming as
\[
\Gt^i{}_j(\Zt^{kl})     &=& 2\,\gd\oversym{ {}^k_j \Zt^{il}} \,, \non[1ex]
\Gt^{ijkl}(\Zt^{mn})    &=& \ft{1}{24}\eps[ijklmnpq]\Zt_{pq}\,,
\lb{27-rep}
\]
where
\[ 
\Zt_{ij} \;:=\; \gOd[ik]\gOd[jl] \Zt^{kl} = (\Zt^{ij})\st\, 
\quad \text{ and } \quad \gOd[ij]\Zt^{ij} \;=\; 0 \,. \nn
\]
Likewise, the \rep{\overline{27}} representation transforms as
\[
\Gt^i{}_j(\Zbt^{kl})    &=& 2\,\gd\oversym{ {}^k_j \Zbt^{il}} \,, \non[1ex]
\Gt^{ijkl}(\Zbt^{mn})   &=& -\ft{1}{24}\eps[ijklmnpq]\Zbt_{pq}\,.
\lb{27b-rep}
\]
Because the product of two $\rep{27}$'s contains no singlet,
there exists no quadratic invariant of $\EEE$; however, there 
is a cubic invariant given by
\[
  \cN_3 (\Zt) := \Zt^{ij} \Zt_{jk} \Zt^{kl} \gOd[il]\,.
\]

We are now ready to give the conformal realization of $\EE$ on the
27 dimensional space spanned by the $\Zt^{ij}$. As the action of 
the linearly realized $\EEE$ subgroup has already been given, we list
only the remaining variations. As before $\Et^{ij}$ acts by
translations:
\[
  \Et^{ij} (\Zt^{kl}) = -\gO[i{[}k]\gO[l{]}j] -\ft18 \gO[ij]\gO[kl]
\]
and $\Ht$ by dilatations
\[
 \Ht (\Zt^{ij} ) = \Zt^{ij} \,.
\]
The $\rep{\overline{27}}$ generators $\Ft^{ij}$ are realized nonlinearly:
\[
\Ft^{ij} (\Zt^{kl}) &:=& -2\,\Zt^{ij}(\Zt^{kl}) 
    + \gO[i{[}k]\gO[l{]}j] (\Zt^{mn}\Zt_{mn})
    +\ft18\, \gO[ij]\gO[kl] (\Zt^{mn}\Zt_{mn})\non[1ex]
&&{}+8\,\Zt\oversym{^{km}\Zt_{mn}\gO[n{[}i] \gO[j{]}l]} 
    - \gO[kl](\Zt^{im}\gOd[mn]\Zt^{nj})
\lb{e7-nonlin-2}
\]
The norm form needed to define the $\EE$ invariant ``light cones''
is now constructed from the cubic invariant of $\EEE$. Then
$\cN_3 (\Xt-\Yt)$ is manifestly invariant under $\EEE$ and under 
the translations $\Et^{ij}$ (observe that there is no need to 
introduce a nonlinear difference unlike for $\E$). Under $\Ht$ it 
transforms by a constant factor, whereas under the action of 
$\Ft^{ij}$ we have
\[
 \Ft^{ij} \Big(\cN_3(\Xt - \Yt) \Big) =
  (\Xt^{ij} + {\Yt}^{ij})  \cN(\Xt - \Yt)\,.
\lb{e7-nonlin-3}
\]
Thus the light cones in $\Rn^{27}$ with base point $\Yt$
\[
\cN_3(\Xt - \Yt) = 0
\]
are indeed invariant under $\EE$. They are still curved hypersurfaces, 
but in contrast to the $\E$ light-cones constructed before, they
are no longer deformed as one varies the base point $\Yt$.

The connection to the Jordan Triple Systems of appendix A can
now be made quite explicit, and the formulas that we arrive 
at in this way are completely analogous to the ones given in the 
introduction. We first of all notice that we can again define a 
triple product in terms of the $\EEE$ representations; it reads
\[
{\JTP{\Xt}{\Yt}{\Zt}}^{ij} 
&=& 
   16\,\Xt\oversym{{}^{ik}\Zti_{kl}\Yt^{lj}}
  +16\,\Zti\oversym{{}^{ik}\Xt_{kl}\Yt^{lj}}
  +4\,\gO[ij](\Xt^{kl}\Yt_{lm}\Zti^{mn}\gOd[kn])\non[1ex]
&&
  +4\,\Xt^{ij}\Yt^{kl}\Zti_{kl}
  +4\,\Yt^{ij}\Xt^{kl}\Zti_{kl}
  +2\,\Zti^{ij}\Xt^{kl}\Yt_{kl}\,.
\]
This triple product can be used to rewrite the conformal realization. 
Recalling that a triple product with identical properties exists
for the 27-dimensional Jordan algebra $J_3^{\On_S}$, we now
now consider $\Zt$ as an element of $J_3^{\On_S}$. Next we
introduce generators labeled by elements of $J_3^{\On_S}$, and 
define the variations
\[
 U_a (\Zt) &=& a \,, \non[1ex]
 S_{ab} (\Zt) &=& \JTP{a}{b}{\Zt} \,, \non[1ex]
\tilde U_c (\Zt) &=& = -\2\JTP{\Zt}{c}{\Zt} \,,
\]
for $a,b,c \in J_3^{\On_S}$. It is straightforward to check
that these reproduce the commutation relations listed in the 
introduction with the only difference that $J_2^\Cn$ has been 
replaced by $J_3^{\On_S}$.

\medskip \noindent {\bf Acknowledgements:} We are very grateful
to R.~ Kallosh for poignant questions and comments on the first
version of this paper. We would also like to thank B.~de Wit 
and B.~Pioline for enlightening discussions.

\newpage
\begin{appendix}
\renewcommand{\theequation}{\Alph{section}.\arabic{equation}}
\renewcommand{\thesection}{Appendix \Alph{section}}

\section{Jordan Triple Systems}\lb{JTS-sec}
\setcounter{equation}{0} 
Let us first recall the defining properties of a Jordan algebra. 
By definition these are algebras equipped with a commutative (but 
non-associative) binary product $a \circ b = b \circ a$ satisfying 
the Jordan identity
\[
(a\circ b)\circ a^2 = a\circ(b\circ a^2) \,. 
\]
A Jordan algebra with such a product defines a so-called Jordan triple 
system (JTS) under the Jordan triple product
\[
 \JTP{a}{b}{c} =  a\circ(\tilde{b}\circ c) +(a\circ \tilde{b})\circ c
 - \tilde{b} \circ(a\circ c) \,, \nn
\]
where {$\; \tilde{ } \;$} denotes a conjugation in $J$ corresponding to 
the operation $\dagger{}$ in \fg. The triple product satisfies the 
identities (which can alternatively be taken as the defining identities 
of the triple system)
\[
\begin{array}{l}
 \JTP{a}{b}{c} = \JTP{c}{b}{a}  \,, \\[1ex]
 \JTP{a}{b}{\JTP{c}{d}{x}}-\JTP{c}{d}{\JTP{a}{b}{x}}
-\JTP{a}{\JTP{d}{c}{b}}{x}+\JTP{\JTP{c}{d}{a}}{b}{x} = 0 \,.
\end{array} \lb{JTP-rel}
\]

The Tits-Kantor-Koecher (TKK) construction~\cite{Tits62,Kant65,Koec67}
associates every JTS with a 3-graded Lie algebra
\[
  \fg = \fg^{-1} \oplus \fg^{0} \oplus\fg^{+1} \,, \label{3-grading}
\]
satsifying the formal commutation relations:
\[
  \Com{\fg^{+1}}{\fg^{-1}} &=& \fg^0 \,, \non
  \Com{\fg^{+1}}{\fg^{+1}} &=& 0     \,, \non
  \Com{\fg^{-1}}{\fg^{-1}} &=& 0     \,. \nn
\]
With the exception of the Lie algebras \Gtwo, \4, and \8 every simple 
Lie algebra $\mathfrak{g}$ can be given a three graded decomposition 
with respect to a subalgebra $\mathfrak{g}^0$ of maximal rank.

By the TKK construction the elements $U_a$ of the $\fg^{+1}$ subspace of the
Lie algebra are labelled by the elements $a\!\in\!J$. Furthermore every such
Lie algebra $\fg$ admits an involutive automorphism$\iota$, which maps the
elements of the grade $+1$ space onto the elements of the subspace of grade
$-1$:
\[
  \iota (U_a) =: \tilde{U}_a \in \fg^{-1} 
\]
To get a complete set of generators of \fg we define 
\[ 
{}[U_a,\tilde{U}_b]   &=& S_{ab} \, ,\non
{}[S_{ab},U_c] &=& U_{\{abc\}} \, ,
\]
where $S_{ab}\in\fg^0$ and $\{abc\}$ is the Jordan triple product under which
the space $J$ is closed. 

The remaining commutation relations are
\[
{}[S_{ab},\tilde{U}_c]    &=& \tilde{U}_{\{bac\}} \, ,\non
{}[S_{ab},S_{cd}] &=& S_{\{abc\}d} - S_{c\{bad\}},
\]
and the closure of the algebra under commutation follows from the defining
identities of a JTS given above. 

The Lie algebra generated by $S_{ab}$ is called the structure algebra of the
$JTS$ $J$, under which the elements of $J$ transform linearly.  The traceless
elements of this action of $S_{ab}$ generate the reduced structure algebra 
of $J$. There exist four infinite families of hermitean JTS's and two 
exceptional ones~\cite{Nehe85,Loo75}. The latter are listed in the
table below (where $M_{1,2}(\On)$ denotes $1 \times 2$ matrices over 
the octonions, i.e. the octonionic plane)
\[
\begin{array}{|c|c|c|} \hline
J & G & H  \\ \hline \vphantom{\int\limits^{a}}
M_{1,2}(\On_S) & \6[(6)]   & SO(5,5) \\[2ex]
M_{1,2}(\On)   & \6[(-14)] & SO(8,2) \\[2ex]
J_{3}^{\On_S}  & \7[(7)]   & \6[(6)] \\[2ex]
J_{3}^{\On}    & \7[(-25)] & E_{6(-26)} \\[1ex] \hline
\end{array}\nn
\]
Here we are mainly interested in the real form $J_{3}^{\On_S}$, which
corresponds to the split octonions $\On_S$ and has \7[(7)] and \6[(6)] as its
conformal and reduced structure group, respectively.

\mathversion{bold}
\section{The quartic \7[(7)] invariant}\lb{Inv-sec}
\mathversion{normal}
\setcounter{equation}{0} 

In the \SLR basis \7[(7)] the quartic invariant is given by~\Eq{e7-invariant},
which we here repeat for convenience
\[
\cI_4^{\scriptscriptstyle{\mathrm{SL}(8,\Rn)}}
&=& X^{ij}X_{jk}X^{kl}X_{li}
    -\ft14 X^{ij}X_{ij}X^{kl}X_{kl} \non
&&
 +\ft{1}{96}\,\eps[ijklmnpq]X_{ij}X_{kl}X_{mn}X_{pq} \non
&&
 +\ft{1}{96}\,\epsd[ijklmnpq]X^{ij}X^{kl}X^{mn}X^{pq} \lb{e7-inv-su} \,.
\]

Another very useful form of $\EE$ makes the maximal compact subgroup \SU
manifest. The fundamental \rep{56} representation then is spanned by the
complex tensors $Z_{AB}$ which are related to the \SLR basis
by~\cite{CreJul79}
\[ 
Z^{AB} \;=\; (Z_{AB})^{*} 
&=& \ft{1}{4\sqrt{2}}(X^{ij} -\i\,X_{ij}) \Gamma^{ij}_{AB}\,,
\] 
where $\Gamma^{ij}_{AB}$ are the \SO gamma matrices. In this basis the quartic
invariant takes the form
\[ 
\cI_4^{\scriptscriptstyle{\mathrm{SU}(8)}} &=& 
   Z^{AB}Z_{BC}Z^{CD}Z_{DA} 
   -\ft14 Z^{AB}Z_{AB}Z^{CD}Z_{CD} \non
&&
 +\ft{1}{96}\,\epsilon^{ABCDEFGH} Z_{AB}Z_{CD}Z_{EF}Z_{GH} \non
&&
 +\ft{1}{96}\,\epsilon_{ABCDEFGH} Z^{AB}Z^{CD}Z^{EF}Z^{GH} \,.
\]
The precise relaton between $\cI_{4}^{\scriptscriptstyle{\mathrm{SU}(8)}}$
and $\cI_{4}^{\scriptscriptstyle{\mathrm{SL}(8,\Rn)}}$ has never been spelled
out in the literature although it is claimed in~\cite{CreJul79} that they
should be proportional. In fact, we have
\[
\cI_4^{\scriptscriptstyle{\mathrm{SU}(8)}} &=& 
 -\cI_4^{\scriptscriptstyle{\mathrm{SL}(8,\Rn)}} \,.
\]
To prove this claim, one needs the identities
\[ 
\mathrm{Tr}(\Gamma^{ij}\Gamma^{kl}\Gamma^{mn}\Gamma^{pq}) &=&
-128\, \delta\undersym{{}^{ij}_{p[k}\delta^{mn}_{l\,]\;q}} 
+128\, \delta\undersym{{}^{ij}_{p[m}\delta^{kl}_{n]q}} 
+128\, \delta\undersym{{}^{ij}_{k[m}\delta^{pq}_{n]l}} \non &&{}
+ 96\,  (\delta^{ij}_{kl}\delta^{mn}_{pq})_{\text{sym}} 
\mp 8\, \epsilon^{ijklmnpq}\,,
\]
and
\[
\epsilon^{ABCDEFGH} \Gamma^{ij}_{AB}\Gamma^{kl}_{CD}
\Gamma^{mn}_{EF}\Gamma^{pq}_{GH} &=& -128\,(
 12\, \delta^{ij}_{kl}\delta^{mn}_{pq}
+48\, \delta\undersym{{}^{ij}_{p[k}\delta^{mn}_{l\,]\;q}})_{\text{sym}} \non &&{}
\mp\,  \epsilon^{ijklmnpq} \,,
\]
where $(\ldots)_{\text{sym}}$ denotes symmetrization w.r.t. the pairs of
indices $(ij)$, $(kl)$, $(mn)$, $(pq)$, and the signs $\mp$ depend on whether
the spinor representation or the conjugate spinor representation of the gamma
matrices is used:
\[
\Gamma^{ijklmnpq} &=& \mp \epsilon^{ijklmnpq} \nn
\]

To see that $\cI_4$ can assume both positive and negative values 
it is sufficient to consider configurations in the $\SU$ basis 
of the form~\cite{FerMal98}
\[
Z_{AB} &=:& \left(
\begin{array}{ccc}
z_{1}&&\\
&\ddots&\\
&&z_{4}
\end{array}
\right)
\otimes
\left( 
\begin{array}{cc}
0&1\\
-1&0
\end{array}\right) \,,
\]
with complex parameters $z_{1},\ldots,z_{4}$.  For this configuration
the quartic invariant becomes 
\[
\cI_4^{\scriptscriptstyle{\mathrm{SU}(8)}} &=&
  \sum_{\alpha} |z_{\alpha}|^{4} 
- 2 \sum_{\beta>\alpha} |z_{\alpha}|^{2}|z_{\beta}|^{2} 
+ 4\, z_{1}z_{2}z_{3}z_{4} + 4 \,z^{*}_{1}z^{*}_{2}z^{*}_{3}z^{*}_{4}\,.
\]
Using this formula, one can easily see that both positive and negative values
are possible for $\cI_{4}$:
\begin{enumerate}
\item [i)] We find positive values for $\cI_{4}$ when all but one parameter
vanish:
\[
\cI_4^{\scriptscriptstyle{\mathrm{SU}(8)}} &=& |z_{1}|^{4} \;>\;0 
 \quad\text{for}\quad z_{1}\neq 0,\, z_{2}=z_{3}=z_{4}=0 \nn 
\]
\item [ii)] $\cI_{4}$ vanishes when all parameters take the same real
(electric) or imaginary (magnetic) value:
\[
\cI_4^{\scriptscriptstyle{\mathrm{SU}(8)}} &=& 0 \quad\text{for}\quad
z_{1}=z_{2}=z_{3}=z_{4}= M \;\text{or}\; \i M\,,\; M\in \Rn  \nn 
\]
This is the example considered in~\cite{KalKol96} corresponding to maximally
BPS black hole solutions in $d\!=\!4$, $N\!=\!8$ supergravity with vanishing
entropy and vanishing area of the horizon.
\item [iii)] $\cI_{4}$ is negative when all parameters take the same
complex ``dyonic'' value. For instance, 
\[
\cI_4^{\scriptscriptstyle{\mathrm{SU}(8)}} &<& 0 \quad\text{for}\quad
z_{1}=z_{2}=z_{3}=z_{4}=\ft{1+\i}{\sqrt{2}}M \,,\;  M\in \Rn \,, \nn 
\]
corresponding to a maximally BPS multiplet with both electric {\em and}
magnetic charges.

\end{enumerate}

\end{appendix}

\end{document}